\documentclass{jpp}
\usepackage{graphicx}

\usepackage[T1]{fontenc}
\usepackage[utf8]{inputenc}
\usepackage{amsmath}
\usepackage{color}
\usepackage{mathtools}
\usepackage{amssymb}
\usepackage{hyperref}
\usepackage{natbib}
\usepackage{subcaption}
\usepackage{booktabs}
\usepackage{caption}

\newcommand{\dd}[1]{\mathrm{d}#1 \,}

\newcommand{\simgt}{\:{\raisebox{-1.5mm}{$\stackrel
{\textstyle{>}}{\sim}$}}\:}

\newcommand{\ee}{\mathrm{e}}

\newcommand{\replace}[1]{}
\newcommand{\zstar}{\rlap{$^{*}$}}
\newcommand{\zminus}{\llap{$-$}}

\shorttitle{\;}
\shortauthor{\;}

\title{Second stability region for gyrokinetics and the L-H transition}

\author{R.J.J.~Mackenbach\aff{1} \corresp{\email{ralf.mackenbach@epfl.ch}}, A.~Zocco\aff{2}, \and P.~Helander\aff{2}}

\affiliation{
\aff{1}École Polytechnique Fédérale de Lausanne (EPFL), Swiss Plasma Center (SPC), CH-1015 Lausanne, Switzerland
\aff{2}Max-Planck-Institut für Plasmaphysik, D-17491 Greifswald, Germany
}

\begin{document}

\maketitle

\begin{abstract}
Using a simple circular tokamak geometry, we show the well-known `second stability region' of MHD-ballooning modes exists for linear gyrokinetics too---whether electrostatic or electromagnetic---and we suggest that the plasma enters this region in H-mode as a consequence of the bootstrap current and Shafranov shift altering the magnetic field, which may occur if the normalised pressure gradient is $\alpha_{\rm MHD} \simgt 1$ and collisionality is low. By performing simulations in more realistic magnetic geometries, we demonstrate a large reduction in collisionless, electrostatic turbulent transport when going from density and temperature profiles typical of L- and H-mode, respectively. This reduction is shown to be a consequence of both the bootstrap current lowering the global magnetic shear, and the pressure gradient altering the local magnetic shear, pushing the plasma towards the second-stability region. A path connecting the L- and H-mode equilibria is constructed, along which the energy and particle fluxes exhibit non-monotonic behaviour as a function of the pressure gradient.
\end{abstract}

\section{Introduction}
Four and a half decades after the discovery of the tokamak H-mode \citep{wagner1982}, no universally accepted explanation has emerged. Enormous progress has been made in the area of gyrokinetic turbulence simulations, which are now able to reproduce the measured transport fluxes and turbulent fluctuations in both L- and H-mode with considerable accuracy, and impressive attempts have been made to simulate the L-H transition itself from first principles \citep{ku2018fast,zholobenko2026fast,de2026lh}. However, to our knowledge, none of these attempts has been independently verified and there is no consensus about the underlying physical mechanism, which is remarkable for a phenomenon that is so robustly attainable in tokamak experiments. It is the aim of this somewhat speculative Letter to propose a reason for this state of affairs and suggest a way forward. 

Most gyrokinetic simulations are conducted in the geometry of a flux tube---a slender volume aligned with the magnetic field that wraps around the torus \citep{beer1995field}. In such simulations, the metric coefficients are constant in the directions perpendicular to the magnetic field but vary along the tube. The unperturbed distribution function is taken to be a Maxwellian with constant density and temperature throughout the tube. Their gradients are also taken to be constant, but the perturbation of the distribution function may vary arbitrarily as long as it satisfies periodic boundary conditions in the perpendicular directions and appropriate (`twist-and-shift') conditions at the ends of the flux tube. In this way, a constant mean gradient of density and temperature is maintained, and the turbulent transport fluxes given these gradients are calculated. 

In an actual tokamak plasma, the equilibrium magnetic field responds to changes in the pressure gradient. If the latter grows, the Shafranov shift increases and the field changes geometry. The local magnetic shear on the outboard side of the torus then tends to decrease, a circumstance that is instrumental in the establishment of the `second stability region' for ballooning modes (\citealt{lortz1978ballooning,coppi1979ideal,freidbergmhd}). In addition, an increased pressure gradient results in more bootstrap current, which also affects the magnetic field and tends to reduce the global magnetic shear in the H-mode pedestal.

These changes to the magnetic shear are known to be stabilising to curvature-driven instabilities, and we suggest that this fact ought to be taken into account in any attempt to simulate the L-H transition. It is not our aim to do so here, but rather to illustrate just how strong the stabilising mechanism is by comparing highly simplified turbulence simulations in magnetic fields corresponding to L- and H-mode, respectively, as will be shown in Section \ref{sec: numerical results}. No attempt is made to make these simulations realistic or to relate them to any specific tokamak experiment---our sole aim is to point out the sheer strength of the stabilising mechanisms and, hopefully, to make a convincing case that the changing magnetic geometry is likely to have a substantial effect on {\em any} turbulence simulation. 

{In the \citet{connor2000review} taxonomy of L-H transition theories, the mechanism this Letter puts forth is among the theories that ``rely on achieving the conditions on plasma parameters for the stabilization of particular instabilities or the suppression of the associated turbulence''. The proposed mechanism is closely related to that of \citet{bishop1986stability,bishop1987bifurcated,bishop1988theory}, who found that sufficient current near the separatrix opens access to a second stability regime for MHD ballooning and interchange modes by lowering the local magnetic shear, yielding bifurcated temperature profiles above a power threshold akin to L- and H-mode. We argue for a similar mechanism, driven here by the bootstrap current and Shafranov shift, producing an analogous second stability region for curvature-driven microinstabilities and a corresponding bifurcation in the turbulent fluxes.\footnote{We note that stabilisation of electrostatic and electromagnetic turbulence due to the Shafranov shift lowering the local magnetic shear has been highlighted before (e.g. in \citealt{bourdelle2003stabilizing,beeke2021impact}), and is expected to play an especially important role in spherical tokamaks \citep{patel2022linear,kennedy2023electromagnetic,giacomin2024electromagnetic}, where the interplay with the ideal-MHD second stability region has been investigated, too \citep{wilson2004integrated,kennedy2026transition}. The role of bootstrap current in the matter is more rarely highlighted, especially in relationship to the L-H transition. Finally, there is rich literature on the physics that limits the H-mode pedestal (examples include \citealt{snyder2002edge,anastopoulos2026impact,10.1063/5.0337081}), and such mechanisms will certainly be important near the end of the transition when the pedestal is nearly established. However, in the earlier stages of the L-H transition, physics that limit the H-mode pedestal are unlikely to be important.} Other notable investigations, which argue for the importance of accounting for an equilibrium-response in the transition, include the works of \cite{Guzdar1996,drake1996local,rogers1998phase}. Finally, these exact mechanisms have been investigated for internal transport barriers, see e.g. \citealt{fukuyama1994theory,bourdelle2005impact,garofalo2015compatibility,staebler2018transport,staebler2018theory,mcclenaghan2019shafranov,ding2024high} and references therein, but are not typically taken into account in modelling the L-H transition itself. Some sophisticated modelling efforts do recognise the variation of the magnetic shear due to the bootstrap current in the pedestal (for example, \citealt{parisi2024stability}), but do not follow it to the conclusion presented here.}

For this investigation, we focus on the simplest possible gyrokinetic turbulence simulations, restricting our attention to electrostatic fluctuations without collisions in Sections \ref{sec: linear analysis}, \ref{sec: nonlinear analysis}, and \ref{sec: bifurcation}. Realistic simulations of the tokamak edge require much more sophistication: magnetic fluctuations \citep{guzdar2001zonal,ashton2025investigation}, sheared rotation \citep{itoh1988model,silva2021structure}, collisions \citep{kerner1998scaling,malkov2015linking}, perhaps global effects \citep{schmitz2017role}, order-unity fluctuations \citep{madsen2013full,michels2021gene}, (neoclassical) effects from large background gradients \citep{nocentini1977transport,Trinczek_Parra_Catto_2025}, finite ion orbit width effects \citep{kiviniemi1998neoclassical,kramer2024formation}, etc., but such simulations are far beyond the scope of this Letter. However, in Section \ref{sec: second-stability} we do show that {\em linear} stability is strongly affected also in the fully electromagnetic case. In the simple $\hat{s}$-$\alpha_{\rm MHD}$ model of tokamak geometry, a second region of stability arises, remarkably similar to that in MHD ballooning-mode theory, when the pressure gradient is large and the magnetic shear is small. Before presenting these results, we explore how the pressure gradient and bootstrap current affect the magnetic geometry, which we do analytically in Section \ref{sec: analytical estimates}, and numerically in Section \ref{sec: equilibria}.  

\section{Analytical estimates} \label{sec: analytical estimates}

We begin by estimating the effect of the bootstrap current on the geometry of the magnetic field in the simple case of a large-aspect-ratio tokamak with circular cross section and small inverse aspect ratio $\epsilon = r/R \ll 1$ (with $r$/$R$ being the minor/major radius). The bootstrap current density is a linear combination of density and temperature gradients, which can be written as 
    \begin{equation}
        j_{\rm BS} = - \frac{qg}{B  \sqrt{\epsilon}} \frac{\mathrm{d}p}{\mathrm{d}r}, 
    \label{JBS}
    \end{equation}
where $B$ denotes the magnetic field strength, $p$ the total plasma pressure, and $q$ the safety factor. The function $g$ depends on the composition of the plasma, the relative length scales of temperature and density variation, and the collisionality $\nu_\ast$ \citep{HS}. In the low-collisionality banana regime, $g$ is of order unity, and becomes much smaller than that at high collisionality. The safety factor is equal to
\begin{equation}
    q(r) = \frac{2B}{\mu_0 \overline{\jmath}(r) R},
\end{equation}
where $\overline{\jmath}(r)$ denotes the average of the plasma current density $j(r)$ over the volume inside the minor radius $r$. The global magnetic shear is defined as
$ \hat{s}(r) = \mathrm{d} \ln q / \mathrm{d} \ln r$ and, in the cylindrical limit, becomes
\begin{equation}
    \hat s(r) = 2 \left( 1 - \frac{j(r)}{\overline{\jmath}(r)} \right). 
    \label{eq:shear-relation}
\end{equation}
In the edge pedestal most of the local current is bootstrap current, and reversing the shear requires $j_{\rm BS} > \overline{\jmath}$
\begin{equation}
    j_{\rm BS}(r) > \frac{2 B}{q\mu_0 R}. 
\end{equation}
In terms of the normalised pressure gradient used in MHD ballooning-mode stability theory, this condition can be expressed as
    \begin{equation} \alpha_{\rm MHD} = - \frac{2 q^2 \mu_0 R}{ B^2} \frac{\mathrm{d}p}{\mathrm{d}r} > \frac{4 \sqrt{\epsilon}}{g}, 
    \label{alpha}
    \end{equation}
where we have used (\ref{JBS}). Formally, the quantity on the right is small, of order $\mathcal{O}(\sqrt{\epsilon})$ in the banana regime, but in practice it is of order unity.

We thus conclude that the bootstrap current makes the magnetic shear negative if (i) the plasma collisionality is sufficiently small and (ii) the pressure gradient is large enough that $\alpha_{\rm MHD} \simgt 1$. If the latter condition is satisfied, the magnetic geometry is also significantly affected by the pressure gradient in the sense that the {\em local} magnetic shear (to be defined in Section \ref{sec: equilibria}) changes appreciably \citep{freidbergmhd}. In addition, the bounce-averaged curvature then becomes favourable for most trapped particles \citep{connor1983effect}. For these reasons, which are all related to the geometry of the magnetic field, curvature-driven gyrokinetic instabilities and the turbulence they drive can be expected to be greatly reduced when $\alpha_{\rm MHD} \simgt \mathcal{O}(1)$. Indeed, in the simple circular-tokamak $\hat{s}$-$\alpha_{\rm MHD}$ model for the magnetic geometry \citep{s-alpha_model},\footnote{It should be noted that this model for the magnetic geometry is formally inconsistent \citep{lapillonne2009clarifications}, but we employ it nonetheless since it is so simple and well known.} we find that gyrokinetic instabilities are suppressed under such conditions (to be discussed in Section \ref{sec: second-stability}). {In diverted plasmas, where the shear diverges towards the separatrix, stabilising effects from bootstrap current are expected to remain, as shown in App. \ref{app: app separatrix}} (though we note that the dependency of the L-H transition on the vertical direction of the $\nabla B$-drift is not due to the mechanisms discussed here). In the next section, we investigate how the bootstrap current and Shafranov shift affect more representative magnetic equilibria.

\section{Numerically computed magnetic equilibria} \label{sec: equilibria}
\begin{figure}
    \centering
    \includegraphics[width=0.8\linewidth]{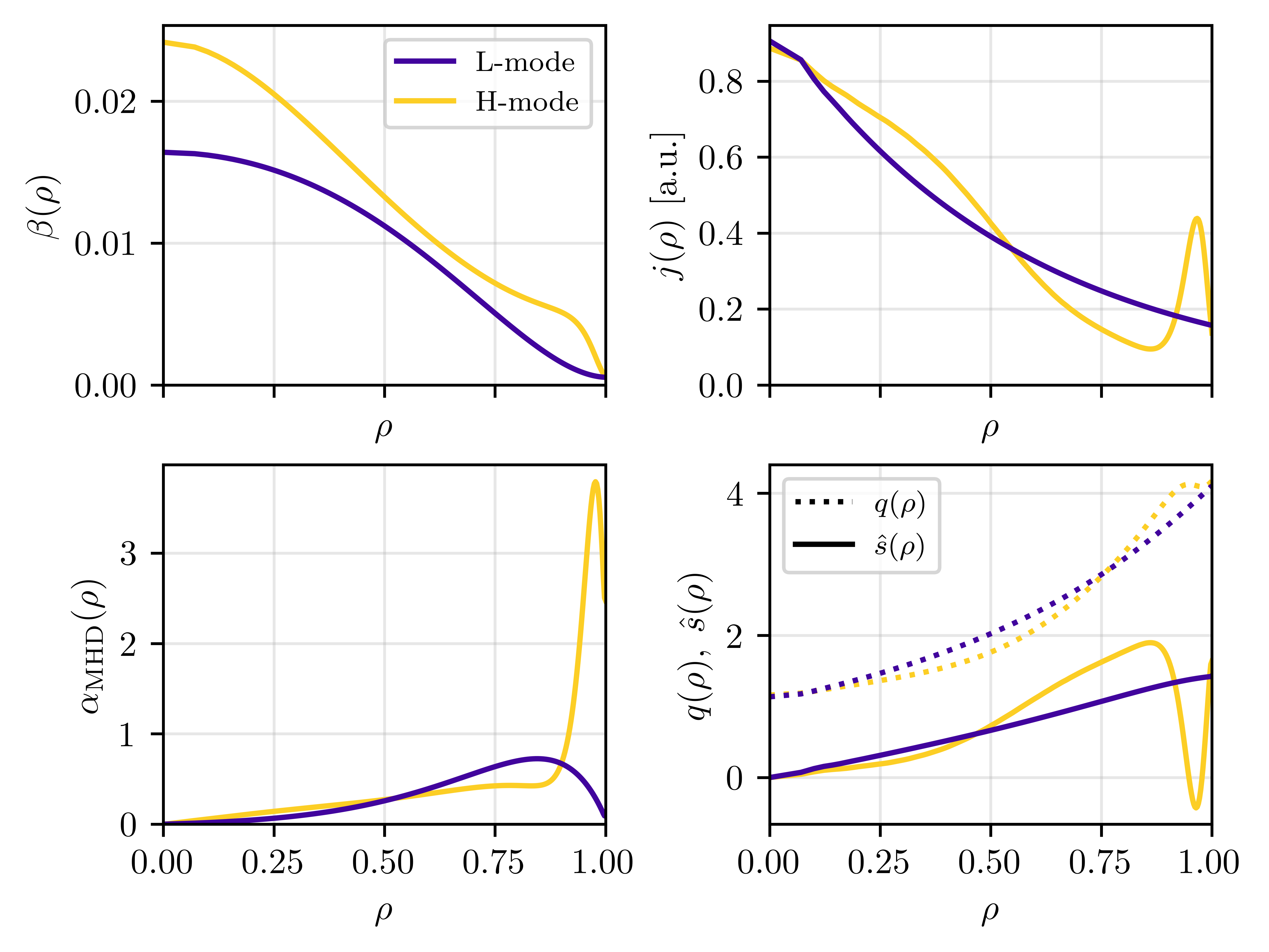}
    \caption{The normalised pressure $\beta=2 \mu_0 p/B_{\rm axis}^2$, current density $j$, normalised pressure gradient $\alpha_{\rm MHD}$, safety factor $q$, and magnetic shear $\hat{s}$ profiles as a function of the radial coordinate $\rho=\sqrt{\psi/\psi_{\rm LCFS}}$.}
    \label{fig: profiles}
\end{figure}
\par 
The shape of the last closed flux-surface, the pressure profile, and the current profile uniquely determine the magnetic equilibrium in an axisymmetric tokamak (if plasma flows are sufficiently small). To isolate the effects described in the introduction, the last closed flux-surface is held constant in all cases considered: it is taken to be a simple vertically elongated plasma whose major axis is a factor $1.35$ larger than its minor axis, and the inverse aspect ratio is $\Delta R/(2R_0) \approx 1/3$ (where $\Delta R$ is the radial extent of the flux-surface). Two pressure profiles are considered: an L-mode profile that gradually decreases radially, and an H-mode profile where the pressure drops sharply near the edge. The first is modelled analytically, using a simple power series for the ion and electron temperatures and density in terms of the radial coordinate. The H-mode profile is numerically interpolated from density and electron temperature measurements in a typical H-mode discharge. 

The \textsc{vmec} code \citep{VMEC} is used in its `fixed boundary mode' to compute the magnetic field. The current profile is described by the radial derivative of the enclosed toroidal current, $\mathrm{d} I_{\rm tor}(\psi)/\mathrm{d}\psi$ ,where $\psi$ is the toroidal flux over $2 \pi$ and serves as a flux-surface label. In Boozer coordinates, this can be expressed as 
\begin{equation} 
    \frac{\mathrm{d}I_{\rm tor}(\psi)}{\mathrm{d}\psi}\propto\frac{q}{qRB_{\varphi}+I_{\rm tor}(\psi)}\left[ \left\langle j_{\parallel}B\right\rangle _{\rm fs}-\frac{\mu_{0}}{2\pi}\frac{\mathrm{d}p}{\mathrm{d}\psi}I_{\rm tor}(\psi)\right] .
    \label{eq: total current}
\end{equation}
where $B_\varphi$ is the toroidal component of the magnetic field, $j_\|$ is the parallel current density, and the flux-surface average is denoted as $\left\langle \dots \right\rangle _{\rm fs}$. The parallel current density is related to gradients in ion temperature ($T_i$), electron temperature ($T_e$), and density ($n$),
\begin{equation}
    \begin{split} \left\langle j_{\parallel}B\right\rangle _{\rm fs}= & \sigma_{\rm neo}\left\langle E_{\parallel}B\right\rangle _{\rm fs} -\\
 & q R B_{\varphi}\left[ p\mathcal{L}_{31}\frac{\mathrm{d}\ln n}{\mathrm{d}\psi}+p_{e}\left(\mathcal{L}_{31}+\mathcal{L}_{32}\right)\frac{\mathrm{d}\ln T_{e}}{\mathrm{d}\psi}+p_{i}\left(\mathcal{L}_{31}+\mathcal{L}_{34}\right)\frac{\mathrm{d}\ln T_{i}}{\mathrm{d}\psi}\right] ,
\end{split}
\end{equation}
so that the total current includes Ohmic, diamagnetic, and bootstrap contributions. Here the neoclassical conductivity is $\sigma_{\rm neo}$, the bootstrap-current coefficients $\mathcal{L}_{kl}$ are complicated formulae given in \citet{Redletal_bootstrap}, $E_{\parallel}$ denotes the parallel inductive electric field, and $p_{i/e}$ is the pressure ions/electrons, respectively. The inductive response can be modelled in a number of ways. Here we aim at studying transport in the region of strong pressure gradients, where most of the current is driven by the pressure gradient (rather than the transformer action). 
We then consider the diamagnetic and bootstrap contributions to dominate, and set $E_{\parallel }=0.$ \par 
The gradient contributions of the current near the edge are calculated by setting the total toroidal current and magnetic flux to $I_{\rm tot}=2$ MA and $\Phi=-12$ Wb, respectively, where the major radius is furthermore set to $R=3$ m, yielding the bootstrap and diamagnetic current-profiles, used to generate  the \textsc{vmec} equilibria.\footnote{In the bootstrap current formulae the trapped particles fraction is approximated in the usual manner $f_{\rm trap}\approx\sqrt{a/R}\approx \sqrt{1/3},$ and the value of $q$ is approximated with its desired edge value. We stress that, consequently, Eq. \eqref{eq: total current} is not satisfied everywhere (one would typically use an iterative scheme to converge to such a global solution), but near the edge region of interest it is nearly.}  
In Figure \ref{fig: profiles}, we show the normalised pressure, current density, $\alpha_{\rm MHD}$, safety factor and magnetic shear profiles of the considered L- and H-mode equilibria.
Its abscissa is $\rho = \sqrt{\psi/\psi_{\rm LCFS}}$, where $\psi_{\rm LCFS}$ is the toroidal flux passing through the last closed flux-surface. It may be seen that the bootstrap current leads to a peak in current density near the edge in H-mode, reducing the safety factor and magnetic shear, where the latter even becomes negative. Around $\rho \approx 0.964$ the shear takes on its most negative value, $\hat{s} \approx -0.4$. Such values are expected to be stabilising for both ballooning modes and micro-turbulence. Further note that, in this region, the Shafranov shift is expected to be especially strong in the H-mode case due to the large pressure gradient, consequently altering the local magnetic shear. \par

At the radial location investigated, the plasma is assumed to consist of singly charged ions (of charge $\ee$) and electrons with density $n_i=n_e=n_0$, $Z_{\rm eff}=1$, and equal temperatures $T_i=T_e=T_0$. The logarithmic pressure gradient thus becomes
\begin{equation}
    \frac{\mathrm{d} \ln p}{\mathrm{d} \rho} = \frac{\mathrm{d} \ln n_0}{\mathrm{d} \rho} \left( 1 + \eta \right),
\end{equation}
where $\eta = \mathrm{d} \ln T_0 / \mathrm{d} \ln n_0$. The L- and H-mode equilibria have different logarithmic pressure gradients, $\mathrm{d}_\rho \ln p \approx -12$ and $\mathrm{d}_\rho \ln p \approx-23$, respectively, but we leave $\eta$ fixed at approximately $1.44$ in both cases. The reason for this choice is to eliminate any effect on the transport that arises from varying $\eta$ (for example, the `stability valley' in maximum-$\mathcal{J}$ devices; see \citealt{alcuson2020suppression}).

It will be of interest to separate the effects of the bootstrap current, which lowers the global magnetic shear, and the Shafranov shift, which modifies the local magnetic shear, to assess the degree to which they affect the transport. For this purpose, two additional equilibria are constructed, where the L- and H-mode current density and pressure profiles are swapped. The equilibrium with L-mode current density and H-mode pressure profile is named $\mathrm{L}_j \mathrm{H}_p$. Conversely, the equilibrium computed using L-mode pressure and H-mode current density is referred to as $\mathrm{L}_p \mathrm{H}_j$. We will thus be simulating turbulence in four different magnetic equilibria: L-mode, $\mathrm{L}_p \mathrm{H}_j$, $\mathrm{L}_j \mathrm{H}_p$, and H-mode. Turbulence in all four of these equilibria is simulated with both the L- and H-mode gradients, thus giving a set of 8 simulations.\footnote{Gyrokinetic simulations can be, and are often, carried out with density and temperature gradients that are not consistent with the pressure profile used in the equilibrium calculation of the magnetic field. Of course, only those simulations that use consistent equilibria are realistic, but inconsistent simulations can be used to gain physical insight, as done here.}  The safety factor $q$ is calculated from the full magnetic equilibrium, and is thus not identical in cases with identical current distributions. However, since $q$ is mostly determined by the total enclosed current, differences are fairly minor at the chosen radial location with its shear being adjusted to $\hat{s} \approx 1.7/-0.7$ in $\mathrm{L}_j \mathrm{H}_p$/$\mathrm{L}_p \mathrm{H}_j$, compared to $\hat{s} \approx 1.3/-0.4$ in L-mode/H-mode.

\begin{figure}
    \centering
    \includegraphics[width=1.0\linewidth]{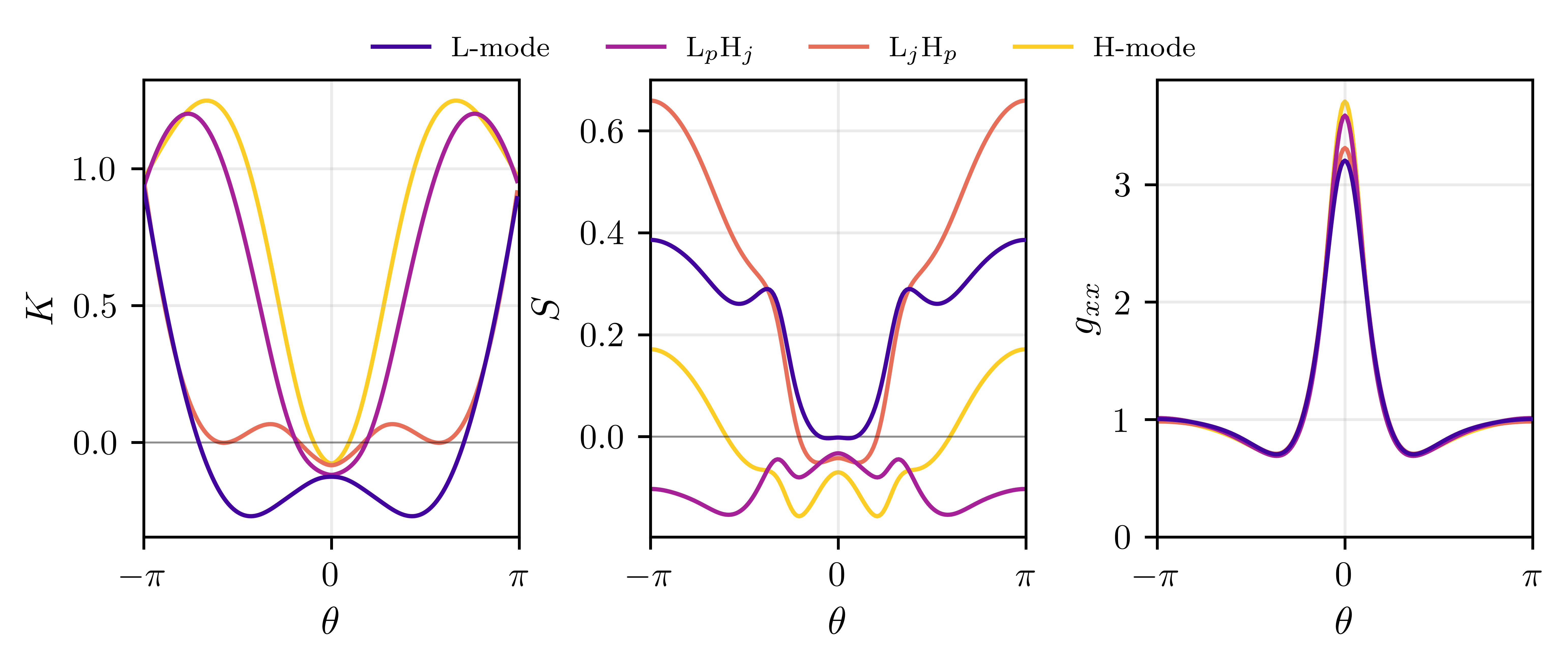}
    \caption{Plots of the gradient drift (left panel), local magnetic shear (central panel), and flux-surface compression (right panel) as a function of the poloidal angle. Negative $K$ corresponds to bad curvature.}
    \label{fig:field-line geometry}
\end{figure}

 In order to assess the geometric differences between the various magnetic equilibria, it is instructive to consider the binormal projection of the gradient drift, the local magnetic shear, and the flux-surface compression as a function of the poloidal angle $\theta$. These are, respectively, defined as
\refstepcounter{equation}
\[
  K = a \frac{\boldsymbol{B} \times \nabla B}{B^2} \cdot \nabla y, \quad
  S = \boldsymbol{X} \cdot (\nabla \times \boldsymbol{X}), \quad g_{xx} = a^2 |\nabla \rho|^2
  \eqno{(\theequation{\mathit{a},\mathit{b},\mathit{c}})}
\]
where $a= \sqrt{2\psi_{\rm LCFS}/B_{\rm ref}}$ denotes the minor radius with $B_{\rm ref}$ being the magnetic field on-axis, $y = \rho a \alpha$ with $\alpha = \theta - \varphi/q$ being the Clebsch angle, and $\boldsymbol{X} = \nabla \alpha - \nabla \psi (\nabla \psi \cdot \nabla \alpha)/(\nabla \psi \cdot \nabla \psi)$ being the component of $\nabla \alpha$ tangent to the flux-surface. The flux-surface average of the local shear is proportional to the global shear \citep{helander2014theory}. Note that negative $K$ corresponds to bad curvature in the chosen coordinate system.

These quantities are displayed in Figure \ref{fig:field-line geometry}, where one sees that L-mode has an especially wide region of `bad' curvature, where $K$ is negative. When either the effect of the larger bootstrap current or that of the steeper pressure profile in H-mode (but not both) is included, the bad-curvature region shrinks, and it becomes yet smaller when both effects are included. The bootstrap current (which is substantial in the $\mathrm{L}_p \mathrm{H}_j$ and H-mode configurations but not in L-mode and $\mathrm{L}_j \mathrm{H}_p$) has a particularly large effect on the overall curvature. The bad-curvature region is particularly small in the H-mode configuration. Next, concerning the local magnetic shear $S$, one observes that it is slightly negative at the outboard mid-plane and largely positive in the bad-curvature region of the L-mode geometry. It becomes more negative in this region when the H-mode pressure profile is adopted ($\mathrm{L}_j \mathrm{H}_p$). If instead the H-mode bootstrap current is adopted ($\mathrm{L}_p \mathrm{H}_j$), the magnetic shear becomes negative everywhere. As a consequence, the global shear then also becomes negative. Finally, the full H-mode geometry exhibits the largest negative local magnetic shear in the bad-curvature region. This property is well known to be helpful for reducing turbulence \citep{Antonsen}. A competing effect of the pressure gradient on curvature-driven instabilities and turbulence is embodied in the flux-surface compression measured by $g_{xx}$, which relates the gradients in flux-space to those in real space, 
\begin{equation}
|\nabla T| = \left|\frac{\sqrt{g_{xx}}}{a} \frac{\mathrm{d}T}{\mathrm{d}\rho} \right|.    
\end{equation}
It essentially reflects the Shafranov shift and thus increases when one goes from L- to H-mode, as can be seen in the right-most panel of Figure \ref{fig:field-line geometry}, though the effect is minor.

\section{Results from gyrokinetic calculations} \label{sec: numerical results}
Our gyrokinetic calculations were performed with the GPU-enabled version of the \textsc{gene} code \citep{jenko2000electron,germaschewski2021toward}. Before presenting the results, we remark that, although these simulations are electrostatic in finite-$\beta$ equilibria and thus incomplete, they retain the effect of the pressure gradient on the particle drift $\boldsymbol{v}_D$. The latter depends on the pressure through the last term in the expression
\begin{equation}
    \boldsymbol{v}_D \propto \boldsymbol{B} \times \left[ \left( \frac{v_\perp^2}{2} + v_\|^2 \right) \frac{\nabla B}{B} + v_\|^2 \frac{\mu_0 \nabla p}{B^2} \right].
\end{equation}
We keep this important effect of the pressure gradient on the particle trajectories,\footnote{In the \textsc{gene} code this is done by setting \texttt{dpdx\_term=`full\_drift'} in the input namelist. The typical convention for simulations without parallel magnetic-field fluctuations is \texttt{dpdx\_term=`gradB\_eq\_curv'}, a choice based on an approximate cancellation in these parallel fluctuations \citep{waltz1999ion,graves2019reduced}, though altering the particle drifts. To verify that this choice is not influential all simulations have been run with the latter choice of \texttt{dpdx\_term} as well. This leads to minor changes of the presented results (notably the nonlinear transport levels in the $\mathrm{L}_j\mathrm{H}_p$ cases are raised by a few tens of percents), but no changes to the conclusions. We finally remark that all nonlinear simulations have been performed with a non-zero but very small $\beta$ ($< 10^{-4}$) to enhance time-stepping, and we have verified that this is inconsequential by running these simulations with $\beta$ halved.} as the magnetic geometry and consequently the trapped-particle precession can be affected by it \citep{rosenbluth1971finite,connor1983effect}. 

\subsection{Second stability region for micro-instabilities} \label{sec: second-stability}
Before considering the full magnetic equilibria computed in Section \ref{sec: equilibria}, we investigate linear gyrokinetic stability in the simple $\hat{s}$-$\alpha_{\rm MHD}$ equilibrium model of a tokamak with circular cross section \citep{s-alpha_model}. This model was originally devised for studying ideal-MHD ballooning modes and led to the discovery of a second stability region, where $\hat s$ is small and $\alpha_{\rm MHD}$ is large. As we shall now see, a similar region exists for gyrokinetic `stability', where the growth rates are much reduced {\footnote{We note in passing that stability in linear gyrokinetics typically refers modes whose growth rate is less than or equal to zero. In this Letter, gyrokinetic `stability' in relation to the $\hat{s}$-$\alpha_{\rm MHD}$ plane refers regions where the growth rates are substantially reduced but not necessarily zero.}} Earlier investigations have, in part, found similar results in specific scenarios or instabilities (see e.g. \citealt{fivaz1997study,joiner2008gyrokinetic,leppin2023complex,parisi2024kinetic} and references therein), but a gyrokinetic stability map in the full $\hat{s}$-$\alpha_{\rm MHD}$ plane in relation to its possible connection to the L-H transition, has not been discussed as far as we are aware.\footnote{The presence or absence of a second stability region for MHD is, however, believed to affect the L-H transition, see e.g. \cite{bishop1987bifurcated,nelson2022prospects}.}

\begin{figure}
    \centering
    \includegraphics[width=1.0\linewidth]{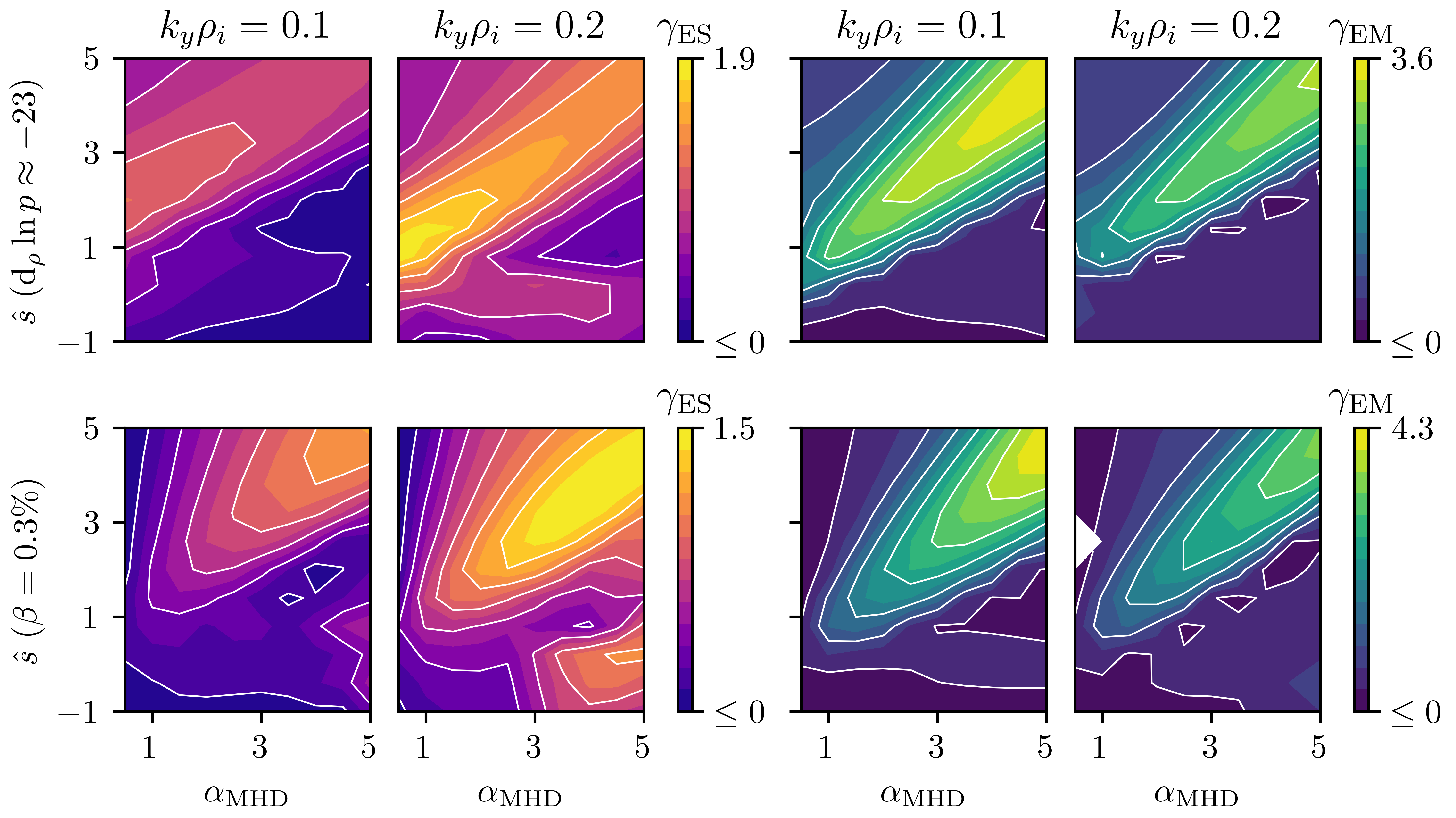}
    \caption{Contour plots of the linear growth rate $\gamma_{\rm ES/EM}$ in units of $a^{-1}\sqrt{T_0/m_i}$, as a function of $\hat{s}$ and $\alpha_{\rm MHD}$. The subscript ES/EM indicates whether the calculation is electrostatic or electromagnetic. Since $\alpha_{\rm MHD} \propto \beta (\mathrm{d}_\rho \ln p) $, the top row has a fixed logarithmic gradient and varying $\beta$, whereas the bottom row has fixed $\beta$ and varying logarithmic gradients. Left-right pairs of panels have $k_y \rho_i = 0.1$ and $k_y \rho_i = 0.2$, respectively, and share the same colour bar to facilitate a comparison. The white regions are calculations that are unconverged due to the very low growth rates, and we note that colourbar maps all stable modes (that is, modes with negative growth rate) onto the same colour as zero.}
    \label{fig:second-stability-region}
\end{figure}

In order to demonstrate gyrokinetic second stability, we explore how the linear growth rate varies with the global magnetic shear and the pressure-gradient parameter (\ref{alpha}) at different fixed wave numbers. Two distinct scans are performed, which differ in how the pressure is treated as $\alpha_{\rm MHD} \propto \beta (\mathrm{d}_\rho \ln p)$ is varied. In the first case, the logarithmic pressure gradient is held constant at the H-mode gradient ($\mathrm{d}_\rho \ln p \approx-23$), a situation that may occur if the pressure itself increases in proportion to its local gradient. In the second case, the local plasma $\beta$ is held fixed at 0.3\%. The remaining parameters are set in close accordance with the geometries detailed in Section \ref{sec: equilibria}; the safety factor is $q\approx4.11$, the aspect ratio is $R_0/a=3$, and the flux-tube is constructed at $r/a = 1$. Finally, we do these scans both electrostatically and electromagnetically, bolstering confidence that our results are not limited to cases without magnetic field fluctuations.
\par 
Contour plots of the linear growth-rate $\gamma$, in units of $a^{-1}\sqrt{T_0/m_i}$ with $m_i$ being the ion mass, are given in Figure \ref{fig:second-stability-region}. The top row has fixed logarithmic gradient (whose contours are broadly in line with results found by \citealt{sona2021electron}), and the bottom panels have a fixed local plasma $\beta$ (whose contours furthermore look like typical ballooning stability diagrams). Furthermore, the binormal wave-number $k_y \rho_i$ of the instability is indicated above each row. Finally, the left panels (with a blue-yellow colourmap) are electrostatic calculations, whereas the right panels (with a green-yellow colourmap) do include magnetic fluctuations, {both parallel and perpendicular}. In all cases one observes two regions with comparatively small growth rates: one with high $\hat{s}$ and low $\alpha_{\rm MHD}$, and one with small $\hat{s}$ and high $\alpha_{\rm MHD}$. The latter may be identified as the second stability region for micro-turbulence, where growth rates (and likely the nonlinear fluxes) are much reduced in spite of the steeper pressure gradient. As discussed in Section \ref{sec: analytical estimates}, it is the effect of the bootstrap current (reducing the magnetic shear) and increasing pressure gradient (raising $\alpha_{\rm MHD}$) that allows the plasma to access this region. Depending on the precise path taken in the $(\hat{s},\alpha_{\rm MHD})$-plane when evolving to the second-stability region, the plasma may become mildly or severely unstable. One can speculate that it should be possible to devise especially favourable paths to the second stability region that circumvent virulently unstable regions, likely reducing the input-power threshold that pushes the plasma into H-mode. We note the contours are similar (although growth rates are typically reduced) for $k_y \rho_i = 0.05$ and $k_y \rho_i = 4.0$. Finally, to investigate robustness we have performed additional scans varying a number of parameters: increasing the ratio of logarithmic gradients $\eta$, using a realistic collisionality, and using realistic geometries. The results from these investigations are presented in App. \ref{app: robustness}. \par 
This idea relies on the notion that the simple $\hat s$-$\alpha_{\rm MHD}$ geometry is sufficiently representative of a true plasma, and that the linear physics are mirrored in the nonlinear transport. To investigate whether this is indeed the case, we turn our attention to the more realistic equilibria discussed in Section \ref{sec: equilibria}, analysing both their linear and nonlinear transport properties.

\subsection{Linear analysis} \label{sec: linear analysis}
\begin{figure}
    \centering
    \includegraphics[width=1.0\linewidth]{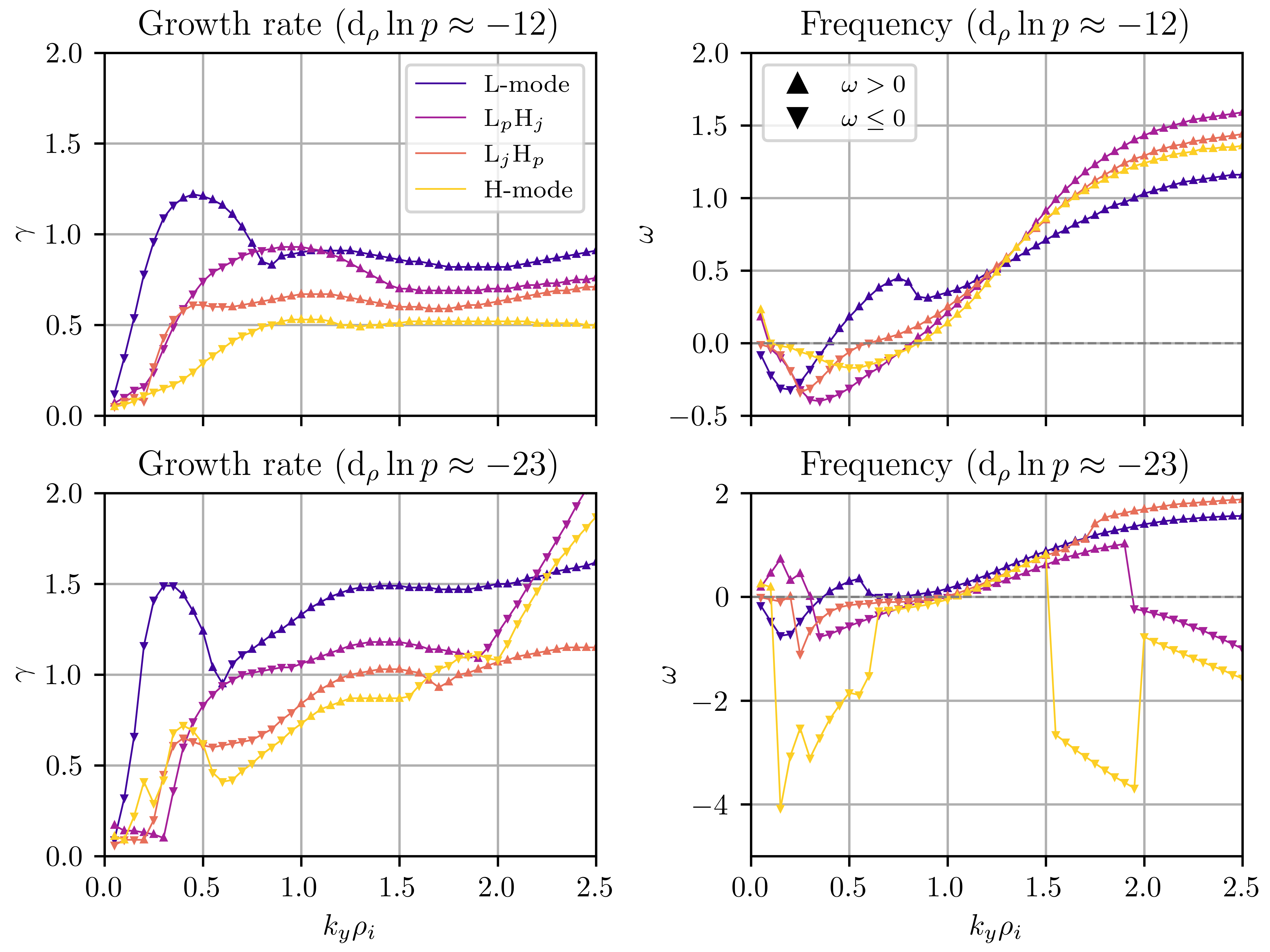}
    \caption{A figure showing the growth rate ($\gamma$) and frequency ($\omega$) of the mode for different binormal wave-numbers $k_y \rho_i$, in units of $a^{-1}\sqrt{T_0/m_i}$. Modes moving in the ion/electron direction have positive/negative frequency and are indicated with triangles pointing upwards/downwards. The colour of the line and marker indicates the magnetic geometry considered, consistently with Figure \ref{fig:field-line geometry}.}
    \label{fig: linear spectra}
\end{figure}
For these equilibria, linear spectra --- growth rates $\gamma$ and frequencies $\omega$ as functions of the binormal wave-number --- are displayed in Figure \ref{fig: linear spectra}, again in units of $a^{-1}\sqrt{T_0/m_i}$ for the equilibria discussed in Section \ref{sec: equilibria}. It may be seen that, in all cases, at low wave-numbers modes rotate predominantly in the electron direction, an indication that the trapped-electron mode is likely the dominant instability here. From a crude `mixing-length' estimate one would expect the instability at these wave-numbers to be the main driver of transport as well. \par
For both sets of logarithmic pressure gradients ($\mathrm{d}_\rho \ln p \approx -12/-23$ for L-mode/H-mode gradients), the growth rate at low wave-numbers is significantly higher in L-mode geometry than in the other geometries. Next, the growth rate in the $\mathrm{L}_p \mathrm{H}_j$ configuration (which includes the H-mode bootstrap current) with L-mode gradients peaks at a higher wave-number, where in the L-mode geometry the peak is significantly lower. In case of H-mode gradients the peaking behaviour is less distinct in $\mathrm{L}_p \mathrm{H}_j$, but the growth rates are lower than in the L-mode geometry. In the $\mathrm{L}_j \mathrm{H}_p$ geometry (which captures the effect of the H-mode Shafranov shift), we see a strong decrease in growth-rate magnitude in both gradients sets. Finally, in the full H-mode case, particularly strong stabilisation can be seen with both gradient sets, where growth rates are reduced by about an order of magnitude at low wave-numbers (and rise to similar values as in the other geometries at high wave-numbers). \par 
We take note of somewhat peculiar behaviour at the lowest wave-numbers in the cases with negative shear (i.e., the $\mathrm{L}_p \mathrm{H}_j$ and H-mode geometries) and H-mode gradients, where growth rates can behave erratically/non-obviously. The mode structure of these instabilities reveals the origin of this behaviour: the mode does not always peak at the central magnetic well and instead establishes at neighbouring magnetic wells in ballooning space. The region of bad curvature in the central magnetic well in cases with negative shear is very narrow and it may be postulated that, e.g., Landau damping stabilises the mode here, a circumstance that is less pronounced for the magnetic well at larger ballooning angles.
\par 
From the linear perspective, these results paint a clear picture in line with Section \ref{sec: second-stability}: there is a stark decrease in instability when going from L- to H-mode, attributed to \emph{both} the decrease in magnetic shear \emph{and} the effects of the Shafranov shift on the geometry. On their own, both are stabilising, but their combined action is much larger: the growth rates drop significantly, especially at low wave-number. As we shall see, this behaviour is mirrored in the nonlinear transport levels, too.

\subsection{Nonlinear analysis} \label{sec: nonlinear analysis}
Nonlinear gyrokinetic simulations were performed of saturated turbulence, taking care to ensure that the simulations were converged and sufficiently well resolved. Energy and particle fluxes were extracted and normalised to gyro-Bohm units defined by
\begin{equation}
  Q_{\rm gB} = T_0 \Gamma_{\rm gB} = \frac{p_0 T_0^{3/2}}{m_i^{3/2} \Omega_i^2 a^2} , \label{eq: gB units}
\end{equation}
where $\Omega_i$ is the ion gyration frequency on the magnetic axis, and $p_0 = n_0 T_0$ is the pressure at the investigated flux-surface. The dimensionless energy and particle fluxes are thus defined as
\refstepcounter{equation}
\[
  Q_s =\frac{1}{Q_{\rm gB}} \left\langle \int \mathrm{d} \boldsymbol{v} \; \frac{m_s v^2}{2} f_{1,s}\boldsymbol{v}_{E} \cdot \nabla \rho \right\rangle , \quad
  \Gamma = \frac{1}{\Gamma_{\rm gB}}  \left\langle  \int \mathrm{d} \boldsymbol{v} \; f_{1,s}\boldsymbol{v}_{E} \cdot \nabla \rho \right\rangle ,
  \eqno{(\theequation{\mathit{a},\mathit{b}})}\label{eq: fluxes definition}
\]
where a subscript $s$ denotes the species label, $f_1$ is the fluctuating part of the distribution function, $\boldsymbol{v}_E$ is the $\boldsymbol{E} \times \boldsymbol{B}$ particle drift, and $ \langle \dots \rangle$ denotes an average over time and the volume of the flux tube. Note that the particle flux of the ions and electrons are identical.
\par
\begin{figure}
    \centering
    \includegraphics[width=1.0\linewidth]{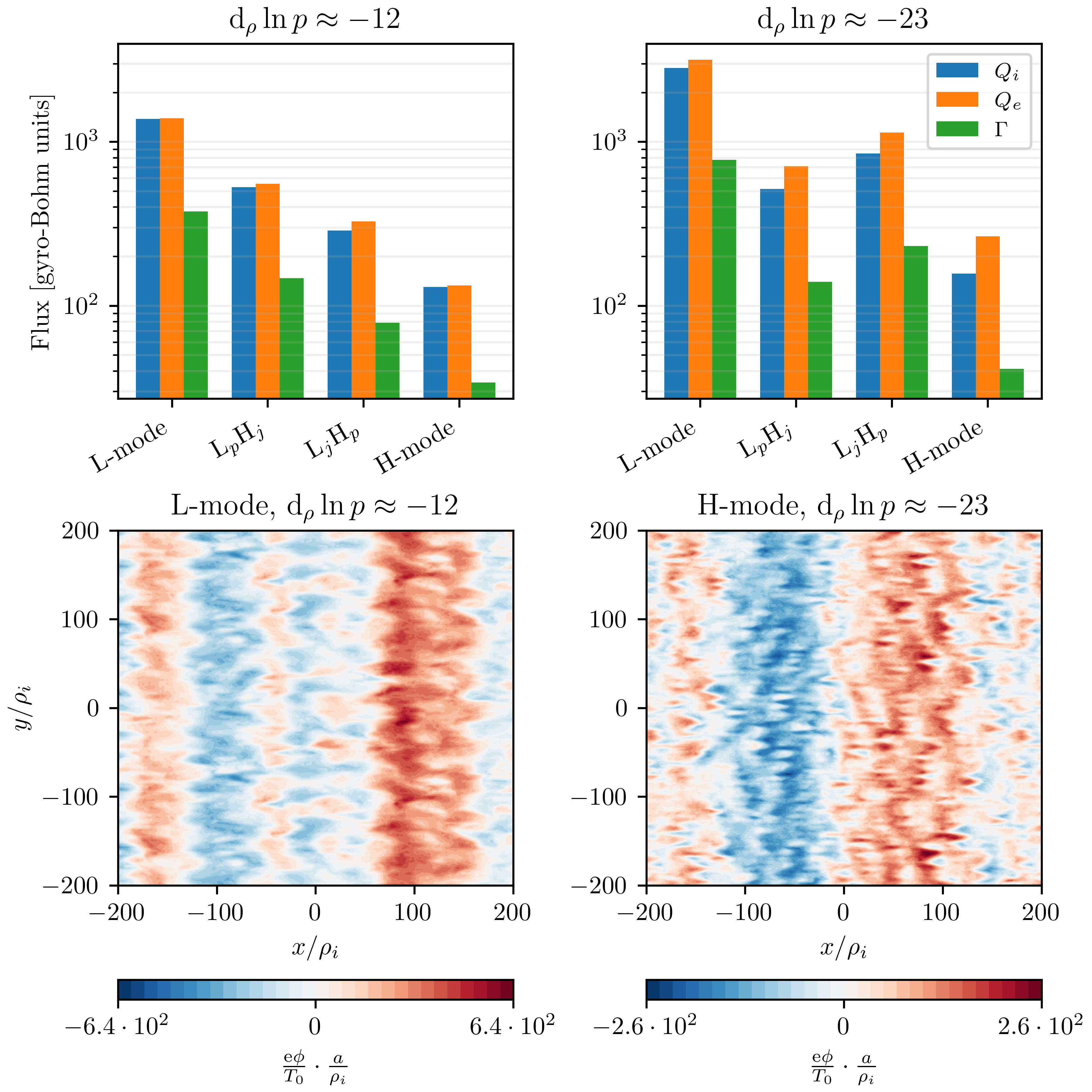}
    \caption{Top row: bar charts of the gyro-Bohm normalised nonlinear fluxes in four geometries, for both L-mode (left) and H-mode (right) gradients. Bottom row: the electrostatic potential ($\phi$, normalised by $T_0 \rho_i a^{-1} \ee^{-1}$) at the outboard-midplane as a function of the radial $(x/\rho_i)$ and binormal $(y/\rho_i)$ coordinates, for the two `consistent' geometries (L/H-mode, with its corresponding gradients).}
    \label{fig:nonlinear fluxes}
\end{figure}
The fluxes are displayed in the top panels of Figure \ref{fig:nonlinear fluxes}, where it may be seen that the fluxes are extremely large, ranging from some tens to thousands of gyro-Bohm units. It is important to realise that the temperature and density are much lower in the edge region, where these simulations are performed, than in the core. As a result, one gyro-Bohm unit of energy flux $Q_{\rm gB} \propto p T^{3/2}$ is a couple of orders of magnitude smaller than in the centre of the plasma. 
\par
More interesting is the massive drop in the fluxes when going from L- to H-mode geometry, where including either the bootstrap current or Shafranov shift reduces the transport somewhat. When both effects are included, however, the drop in transport is far more pronounced. At fixed values of the density and temperature gradients, this reduction is well above a factor ten, with the largest decrease being close to a factor twenty. Note that only accounting for the Shafranov shift when adjusting the equilibrium (i.e., comparing the L-mode case with $\mathrm{d}_\rho \ln p \approx -12$ with the $\mathrm{L}_{j} \mathrm{H}_p$ case with $\mathrm{d}_\rho \ln p \approx -23$), it can be seen that one could support roughly twice as steep logarithmic gradients for the same flux in gyro-Bohm units, but, when accounting for the bootstrap current also, one can support much steeper gradients. Finally, the $\mathrm{L}_p\mathrm{H}_j$ case exhibits surprising behaviour; the fluxes remain roughly the same when the logarithmic gradients are nearly doubled. This seems to be a consequence of a shift in the dominant wave-number in the turbulent spectrum---with L-mode gradients the peak is at $k_y \rho_i \approx 0.03$, whereas in the H-mode case the peak is nearly an order of magnitude higher at $k_y \rho_i \approx 0.25$. \par

It is important to observe that the two `consistent' geometries (L- and H-mode with its corresponding gradient set) have different transport levels too, where L-mode suffers from much higher gyro-Bohm normalised fluxes than H-mode. As can be seen in the bottom panels of Figure \ref{fig:nonlinear fluxes}, this reduction in transport in H-mode is likely a consequence of the smaller eddy sizes and free energy in the H-mode geometry, as compared to L-mode. This difference in transport of the consistent equilibria has an important corollary, namely that the fluxes must be non-monotonic functions of the gradients (since in the limit of zero gradients there is no transport). With this type of turbulence, there is a bifurcation in the solution to the transport equations: for a certain range of fluxes, there are two solutions, corresponding to small and large gradients, respectively.\footnote{There is experimental evidence for such bifurcations in the context of internal transport barriers; see Figure 3 of \citet{mcclenaghan2019shafranov}.} 
\subsection{Bifurcation} \label{sec: bifurcation}
\begin{figure}
    \centering
    \includegraphics[width=1.0\linewidth]{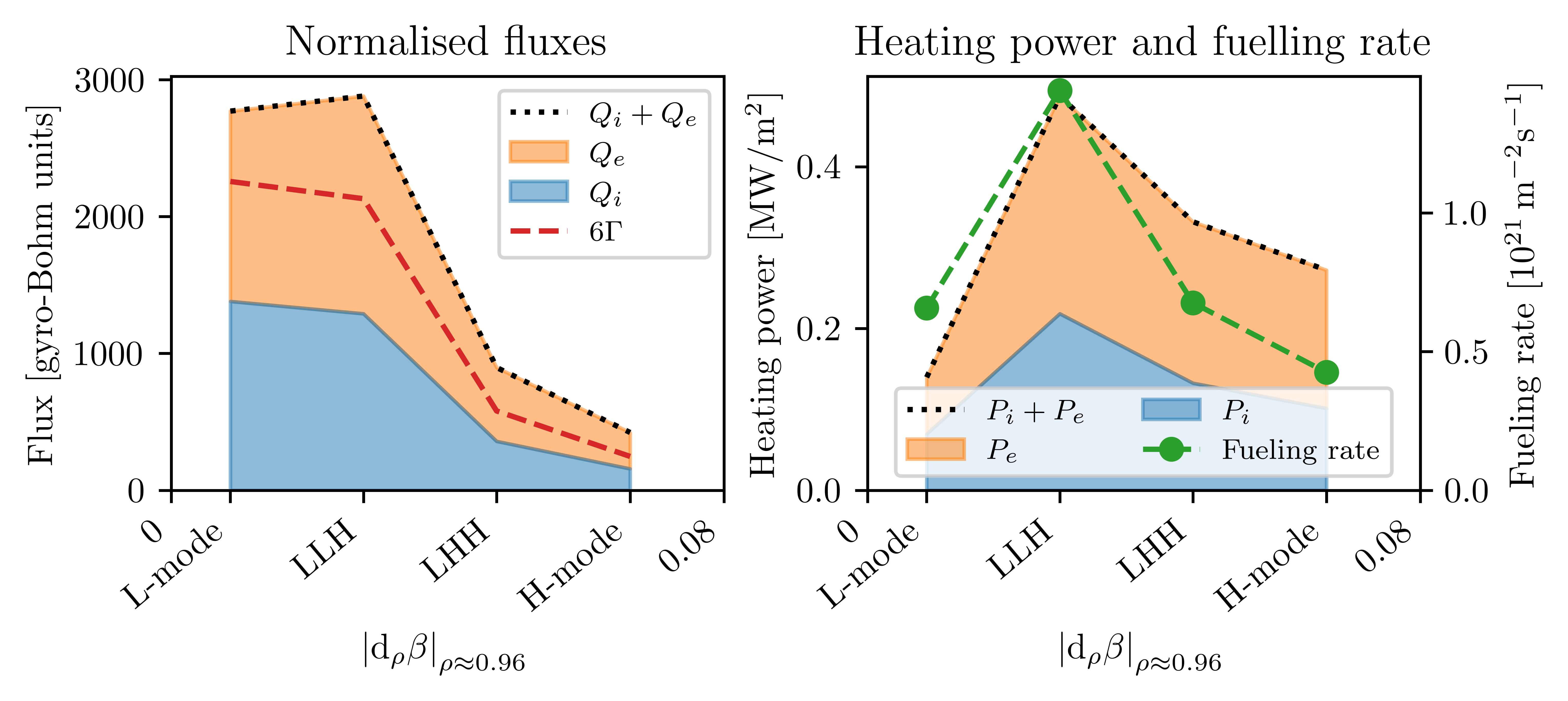}
    \caption{Left: the gyro-Bohm normalised fluxes as a function of the real-space pressure gradient, in the four geometries considered. Right: the input power and fuelling rate per square meter (accounting for the changing gyro-Bohm unit). The particle fluxes have been multiplied by a factor 6 in the left panel to make its trends more discernable. Finally the abscissa goes from $|\mathrm{d}_\rho \beta|\in[0,0.08]$, with the gradient of the various geometries labelled by the geometry name.}
    \label{fig:nonlinear fluxes bifurcation}
\end{figure}
To investigate this bifurcation, two interpolated equilibria that lie between the L- and H-mode fields are constructed. Since the relationship between the bootstrap current and the equilibrium profiles is not obvious when evolving from L- to H-mode, we opt to use the simple linear interpolation of the pressure and current density profiles. In reality details will depend on the heating and fuelling profiles as well as their time history, but it is not our aim to dwell on such details, but rather to show that such a bifurcation exists for a smoothly varying sequence of equilibria. (In the logic of Section \ref{sec: second-stability}, we are choosing a path in an $\hat{s}$-$\alpha_{\rm MHD}$ parameter space and the transport along this path is investigated.) To this end two intermediate equilibria have been constructed: LLH/LHH, where the pressure and current profile are the average of two/one L-mode and one/two H-mode equilibria. Flux tubes are generated at $\rho\approx0.964$, and, for the progression $[\text{L-mode},\text{LLH},\text{LHH},\text{H-mode}]$, the logarithmic pressure gradient here varies as $\mathrm{d}_\rho \ln p \approx [-12,-19,-21,-23]$ whereas the global magnetic shear goes as $\hat{s} \approx [1.4,0.8,0.2,-0.4]$. The binormal drift, local magnetic shear, and flux-surface compression are given in Appendix \ref{app: linear spectra} (where a linear analysis is furthermore included), and we simulate the turbulence in these flux-tubes setting $T_e=T_i$ and $\eta \approx 1.44$, as before. \par 
The resulting transport, viz. the gyro-Bohm normalised energy and particle fluxes, are displayed in the left panel of Figure \ref{fig:nonlinear fluxes bifurcation}, where one observes very slight non-monotonic behaviour in the energy fluxes. However, one should keep in mind that the gyro-Bohm unit also varies when the equilibrium is changed. To this end, we have converted these fluxes into input powers and fuelling rates per unit area (in accordance with Eq. \ref{eq: gB units}). {It should be noted that the density and temperature of the interpolated equilibria are underdetermined: since we have only specified the pressure, this fixes the product but not the individual values. To this end we have interpolated both geometrically, i.e. $h_{\rm interp}=h_L^{1-x}h_H^{x}$, with $h$ being density/temperature, choosing $x$ such that the pressure is matched.} The input power and fuelling rate per unit area are displayed in the right panel of Figure \ref{fig:nonlinear fluxes bifurcation}, where the non-monotonic behaviour is far more pronounced and one may observe that for a certain range of the inputs powers and fuelling rates there are two corresponding pressure gradients: a shallow and steep gradient solution. One can furthermore speculate that the maximum value in this Figure is the approximate flux required to access the steep-gradient solution, and such power-thresholds are indeed observed \citep{Y_R_Martin_2008}. Approximating the flux-surface area as $(2\pi)^2R_0a\approx118 \; {\rm m^2}$ the maximal flux becomes about 60 MW, somewhat higher than the typical $\sim$10 MW, which may be a consequence of the various simplifications made.

\section{Conclusions \& discussion}
In this Letter, we have explored the importance of varying the magnetic geometry consistently with the evolution of the pressure and current profiles in simulations of turbulence and transport in tokamaks. As the pressure gradient increases, the magnetic shear and the spatial distribution of `bad' curvature are modified both by the deformation of the flux-surfaces and by the increasing bootstrap current. These changes in the magnetic geometry push the plasma toward the second stability region of MHD ballooning modes. A similar region at large $\alpha_{\rm MHD}$ and small $\hat{s}$ has been found for gyrokinetic instabilities, both electrostatic and electromagnetic ones, whose linear growth rates are greatly reduced, at least in the case of simple circular geometry.
\par
Linear stability calculations of electrostatic instabilities were performed in more realistic equilibria too, showing a similar reduction in growth rates when going from typical L- to H-mode plasma profiles. The stabilisation occurs thanks to changes in the magnetic geometry caused by the bootstrap current and the deformation of the flux surfaces associated with the strong pressure gradient. This behaviour is shown to be mirrored in nonlinear turbulence simulations as well, where the transport levels decrease by well over an order of magnitude. Finally, these stabilising effects result in non-monotonic behaviour of the fluxes: in a certain range of gradients typical of the plasma edge, the transport decreases with increasing density and temperature gradients. For simplicity, our simulations are local, electrostatic and collisionless, but the reduction in transport is so large that it would be surprising if it does not carry over to more complicated situations. \par
It is tempting to speculate that the H-mode arises, at least in part, because of the mechanisms we have described: if the edge plasma is sufficiently collisionless, an increasing pressure gradient will drive a substantial bootstrap current, which reduces the magnetic shear (in addition to the direct effect of the Shafranov shift) and stabilises micro-instabilities. Much more detailed simulations would be needed to confirm this picture, but we note that the reason why no such first-principles simulations have uncovered these results before may be that the magnetic equilibrium is rarely evolved consistently with the plasma profiles.\footnote{It should be noted that in the integrated and reduced modelling communities, the important role of consistent magnetic equilibria has been recognised for some time for steady-state prediction profiles (e.g. \citealt{meneghini2016integrated,lyons2023flexible}). Seeing its importance in that field, it stands to reason that a dynamical simulation of the transition should take it into account equally.} This requires coupling the micro-turbulence, neoclassical, and transport time-scales, a computationally demanding endeavour. A code-suite like \textsc{gene--tango}, \textsc{gx--trinity}, or \textsc{cgyro--portals} may be especially well-suited for this task \citep{di2022global,qian2022stellarator,rodriguez2024enhancing}, if the radial resolution is sufficient and, crucially, the magnetic equilibrium is continually updated as the simulation proceeds. Alternatively, the simulation could be carried out using a novel global gyrokinetic code that accounts for electromagnetic and neoclassical effects, in particular the bootstrap current, and the effect of the latter on the magnetic geometry. \par 
{Finally, we note that such detailed simulations may uncover under what conditions the described mechanisms are most prominent---we have highlighted the importance of having sufficiently small collisionality, but it may furthermore be important to have a conditioned wall \citep{kaye2011h}, flux-surfaces of certain shapes \citep{davies2022kinetic}, a strong enough density gradient \citep{kotschenreuther2024transport}, certain thermodynamic conditions \citep{mahajan2026unified}, etc. We leave such questions open to future investigations.}

\section*{Acknowledgements}
It is a pleasure to thank Justin Ball, Stephan Brunner, Stefano Coda, Baptiste Frei, Joaquim Loizu, Gareth Roberg-Clark, Eduardo Rodr\'iguez, Fritz Wagner, and Wladimir Zholobenko for valuable discussions. This publication is part of the project `Enabling a star on Earth with thermodynamics: A path to viable fusion power plants' with project number \texttt{019.241EN.015} of the research programme \emph{Rubicon} which is (partly) financed by the Dutch Research Council (NWO). This work was supported by a grant from the Swiss National Supercomputing Centre (CSCS) under project ID \texttt{lp72} and \texttt{lp131} on Alps--Daint. RJJM furthermore gratefully acknowledges the computing time provided by the Max Planck Computing and Data Facility on the HPC systems Raven and Viper.

\section*{Data availability}
Data of the magnetic equilibria, the magnetic field lines, and scripts that generate the Figures, are accessible via a Zenodo archive with DOI: \texttt{10.5281/zenodo.23943351}. Condensed data of the gyrokinetic simulations are included here as well. The full data of the simulations can be made available upon reasonable request.

\appendix

\section{Robustness of the second stability region} \label{app: robustness}
{In Fig. \ref{fig:second-stability-region} one observes a second-stability region for collisionless linear gyrokinetics (where, strictly speaking, growth rates do not vanish but are much reduced) in $\hat{s}$-$\alpha_{\rm MHD}$ geometry, with the logarithmic ratio of gradients set to $\eta\approx 1.44$. In this appendix, we explore how sensitive this region is to various assumptions, by performing the same scan with (i) doubled $\eta$, (ii) increased collisionality, and (iii) realistic magnetic geometries. \par 
To compute the latter, \textsc{vmec} equilibria are constructed whose last closed flux-surface has a shape as parametrised by \citet{miller_noncircular_1998}, with an elongation and triangularity of $\kappa=1.5$ and $\delta = \pm 0.3$. Subsequently choosing the pressure and current profile typical of L-mode/H-mode and running \textsc{vmec}, one may extract the remaining Miller parameters at $\rho\approx0.964$, which are then guaranteed to correspond to a global equilibrium. Doing so for L- and H-mode yields nearly identical parameters, \emph{except} for the values of $\hat{s}$ and $\alpha_{\rm MHD}$. For the stability diagrams, the latter were varied whilst the former Miller parameters were held constant at the values given in Table \ref{tab:tri} for positive and negative triangularity, respectively. 
 \par
\begin{table}
  \begin{center}
\def~{\hphantom{-}}
  \begin{tabular}{lcccccccc}
      Geometry & $R_0\; [{\rm m}]$ & $r/R_0$ & $q_0$ & $\kappa$ & $s_\kappa$ & $\delta$ & $s_\delta$ & $\partial_r R$ \\
      Pos. tri. & 3.007 & 0.323 & 3.874 & 1.486 & 0.306 & ~~0.284 & ~~0.477 & ~~\zminus$0.182$ \\
      Neg. tri. & 3.012 & 0.321 & 3.907 & 1.498 & 0.099 & ~~\zminus$0.288$ & ~~\zminus$0.391$ & ~~\zminus$0.331$ \\
  \end{tabular}
  \caption{Miller parameters of the positive- and negative-triangularity equilibria considered in this Appendix, definitions of quantities as in \citet{miller_noncircular_1998}}
  \label{tab:tri}
  \end{center}
\end{table}
The resulting stability maps are depicted in Fig. \ref{fig:robustness}, where the second stability region clearly persists, although its shape and the overall magnitude of the growth rates vary. We thus conclude that the existence of the second-stability region is robust. 
\begin{figure}
    \centering
    \includegraphics[width=1\linewidth]{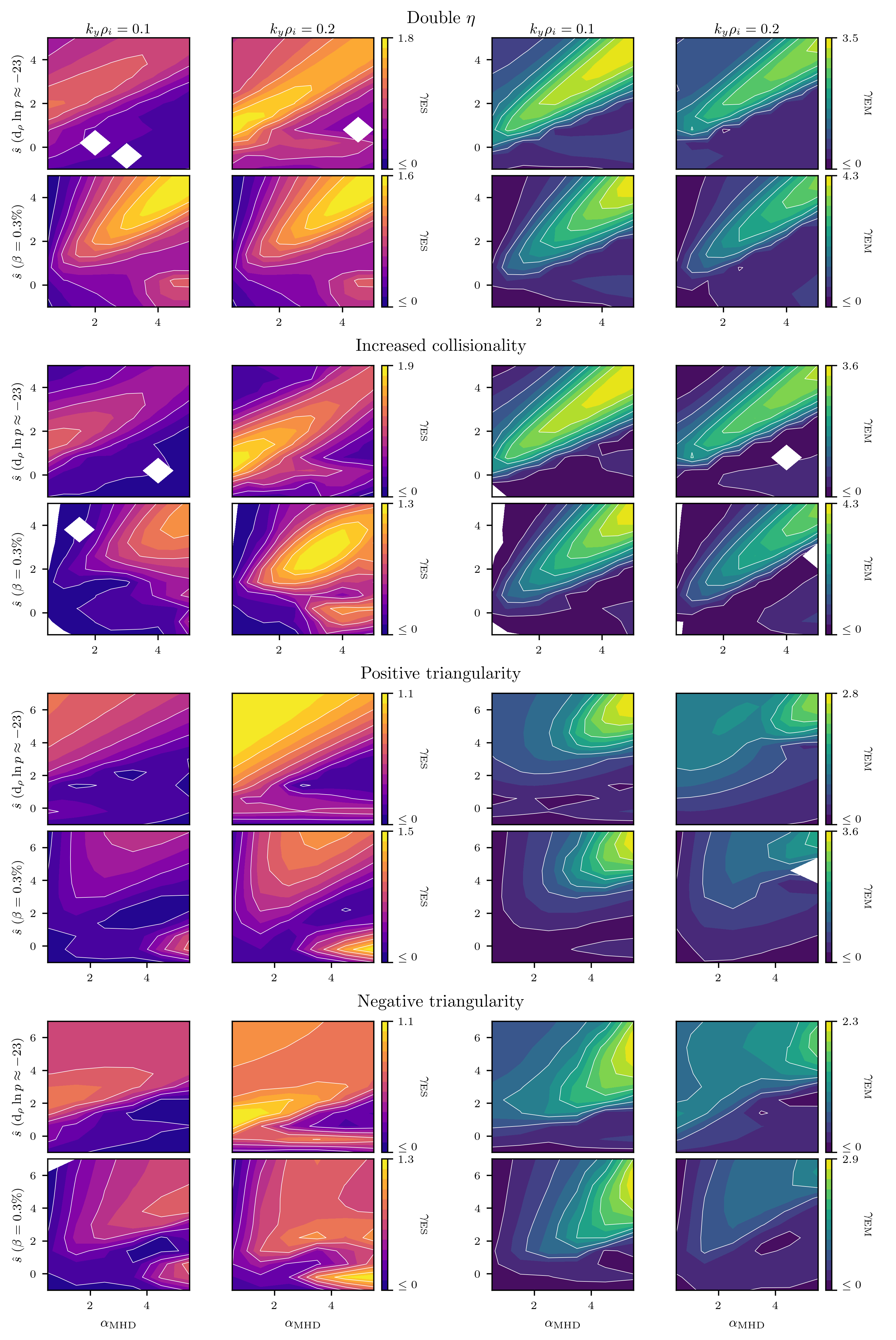}
    \caption{Scans over $\hat{s}$-$\alpha_{\rm MHD}$ as in Fig. 3, but with various adjustments: the first (top) set of eight panels has $\eta\approx2.88$, the second set of eight panels has collisionality (in \textsc{gene} units) of $\nu_*=0.01$, the third (positive triangularity) and fourth (negative triangularity) set of eight panels are performed in the realistic geometries given in Table \ref{tab:tri}.}
    \label{fig:robustness}
\end{figure}

}

\section{Additional details of the interpolated equilibria, and simulations} \label{app: linear spectra}
\begin{figure}
    \centering
    \includegraphics[width=1.0\linewidth]{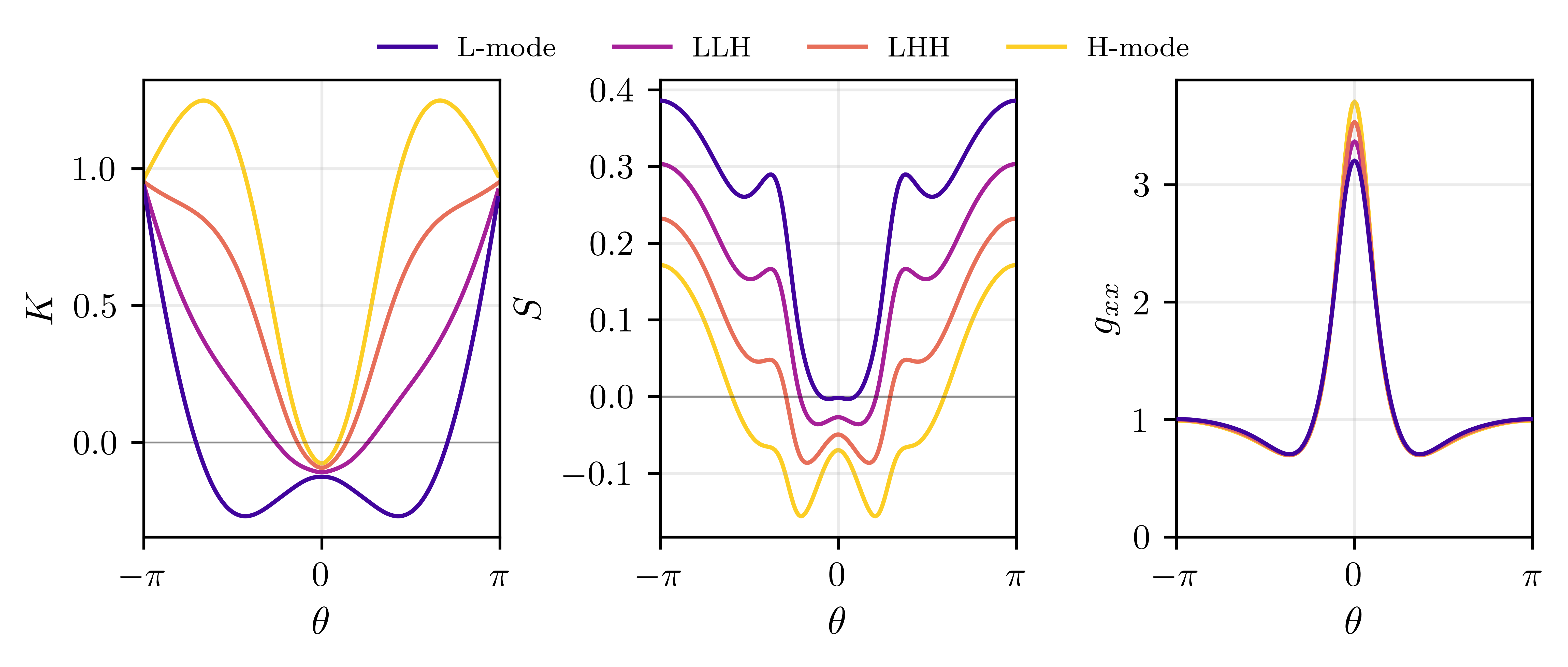}
    \caption{Plots of the gradient drift (left panel), local magnetic shear (central panel), and flux-surface compression (right panel) as a function of the poloidal angle, for the interpolated equilibria.}
    \label{fig:field-line geometry interp}
\end{figure}
\begin{figure}
    \centering
    \includegraphics[width=1.0\linewidth]{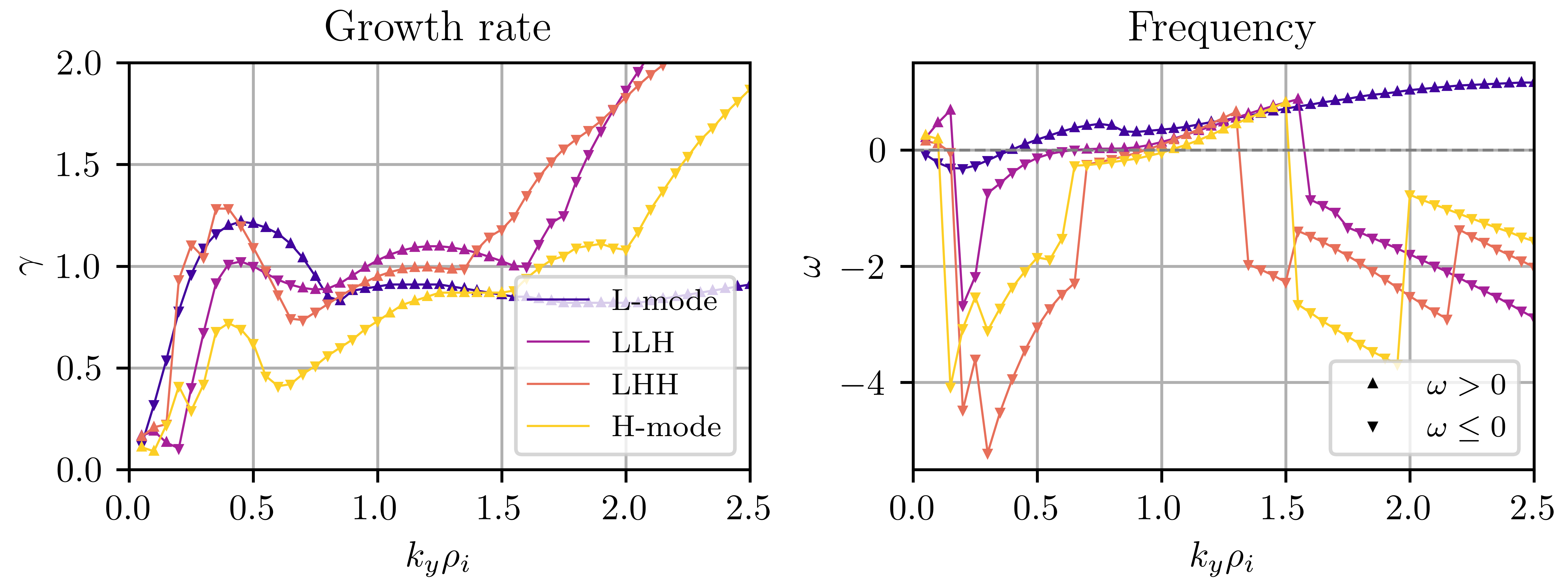}
    \caption{The linear spectra in the interpolated equilibria. The colour of the line and marker indicates the magnetic geometry considered, consistently with Figure \ref{fig:field-line geometry interp}.}
    \label{fig:linear spectra interp}
\end{figure}

In this appendix additional details of the interpolated magnetic equilibria and simulations are given. Similarly to Figure \ref{fig:field-line geometry} the magnetic field-line geometry of the L-mode, LLH, LHH, and H-mode equilibria are displayed in Figure \ref{fig:field-line geometry interp}. As the magnetic geometry goes from L- to H-mode, it may be seen that the region of bad curvature shrinks, the local magnetic shear in this region becomes more negative, and the flux-surface compression at the outboard-midplane increases slightly. The linear spectra in these equilibria is given in Figure \ref{fig:linear spectra interp}. It is clear that the overall growth rate reduces and the spectrum peaks at higher wave-numbers in LLH as compared to L-mode. However, the growth-rate in LHH increases somewhat again, further peaking at lower wave-number. Qualitatively, this does not correspond to the behaviour of the nonlinear transport in Figure \ref{fig:nonlinear fluxes bifurcation}, hinting at non-obvious nonlinear effects as one moves from L- to H-mode in this scenario. \par
{For completeness, we also list the device parameters of the $\hat{s}$-$\alpha_{\rm MHD}$ and all \textsc{vmec} equilibria (L-mode, H-mode, $\mathrm{L}_p \mathrm{H}_j$, $\mathrm{L}_j \mathrm{H}_p$, LLH, and LHH), as well as the local parameters of the simulated flux-tube, in Table \ref{tab:vals}. Several quantities are shown: $B_{\rm axis}$ is the reference magnetic field on axis, $R_{\rm axis}$ is the major radius of the axis, $a_{\rm minor}$ is the minor radius as defined by $(R_{\rm max}-R_{\rm min})/2$, $\kappa_{\rm LCFS}$ is the elongation of the last-closed flux surface (the vertical over the radial extent), $\rho_{\rm ft}$ is the radial position at which the flux-tube is constructed, and the normalised gyroradius and collision frequency are defined as in \textsc{gene} $\rho_*=\sqrt{m T_0}/\ee B_{\rm axis}$ and $\nu_*= \pi \ln\Lambda \ee^4 n_0 a_{\rm minor}/(2^{3/2} T_0^2)$ (in Gaussian units). We note that in the $\hat{s}$-$\alpha_{\rm MHD}$ equilibrium many parameters are either not defined, since certain dimensions are normalised out, or scanned over. Importantly, the density and temperature in the interpolated equilibria are underdetermined, since only the pressure is being interpolated leaving some freedom in $n_0$ and $T_0$. Treating density and temperature on equal footing, a natural parametrisation is $n_{\rm interp}=n_L^{1-x}n_H^{x}$ and $T_{\rm interp}=T_L^{1-x}T_H^{x}$ where $x$ is chosen so that the pressure is matched, which results in the values given in the Table. These interpolated values and derived quantities are indicated with a superscript $*$. The collisionality $\nu_*$ (in \textsc{gene} units, as given in the input namelist) and plasma $\beta$ are included here for completeness: we stress that all presented simulations (apart from the electromagnetic $\hat{s}$-$\alpha_{\rm MHD}$ scans and collisional scan in App. \ref{app: robustness}) are collisionless and electrostatic. \\
Finally, we include details of convergence checks performed: for all simulations presented in this Letter we have (i) doubled the parallel resolution, (ii) doubled the radial resolution, (iii) doubled the perpendicular-velocity resolution, (iv) doubled parallel-velocity resolution, (v) doubled the extent of the simulation box in the perpendicular-velocity direction at double resolution, (vi) doubled the extent of the simulation box in the parallel-velocity direction at double resolution, (vii) halved the hyperdiffusion in the parallel direction, (viii) halved the hyperdiffusion in the parallel velocity direction, and (ix) changed \texttt{dpdx\_term} from \texttt{`full\_drift'} to \texttt{`gradB\_eq\_curv'}. Nonlinear simulations have additionally been checked for convergence by (i) doubling the number of Fourier modes in the binormal direction, (ii) doubling the binormal box-size at double resolution, (iii) doubling the radial box-size at double resolution, (iv) doubling the simulation time, (v) halving $\beta$, verifying if quantities of interest do not change significantly. The exact parameters per simulation can be found in the Zenodo archive. We note that only the fixed-$\beta$ electrostatic panels of the double $\eta$ case of Fig. \ref{fig:robustness} have not been checked for convergence for a lack of computational time.
\begin{table}
  \begin{center}
\def~{\hphantom{-}}
  \begin{tabular}{lcccccccc}
      Quantity & Unit  &   $\hat{s}$-$\alpha_{\rm MHD}$    &   L-mode  &   H-mode  &   $\mathrm{L}_p \mathrm{H}_j$ &   $\mathrm{L}_j \mathrm{H}_p$ &   LLH &   LHH \\[3pt]
      $B_{\rm axis}$ & $[{\rm T}]$                 &   N/A &   2.67&   2.67&   2.67&   2.67&   2.67&   2.67\\
      $R_{\rm axis}$& $[{\rm m}]$        &   3.00 &   3.00 &   3.00 &   3.00 &   3.00 &   3.00 &   3.00 \\
      $a_{\rm minor}$& $[{\rm m}]$       &   1.00 &   1.00 &   1.00 &   1.00 &   1.00 &   1.00 &   1.00 \\
      $\kappa_{\rm LCFS}$&   &   1.00 &   1.35&   1.35&   1.35&   1.35&   1.35&   1.35\\
      $\rho_{\rm ft}$    &   &   1.00 &   0.96&   0.96&   0.96&   0.96&   0.96&   0.96\\
      $q(\rho_{\rm ft})$ &   &   4.11&   3.89&   4.11&   4.09&   3.91&   3.97&   4.04\\
      $\alpha_{\rm MHD}(\rho_{\rm ft})$&&N/A&0.39&3.37&  0.43&   3.05&   1.32&   2.31\\
      $\hat{s}(\rho_{\rm ft})$& & N/A & 1.39 & \zminus0.43 & \zminus0.72 & 1.67 & 0.77 & 0.16\\
      $n_0(\rho_{\rm ft})$& $[10^{19}\,{\rm m}^{-3}]$    &   N/A& $3.57$ & $6.61$ & $3.57$ &  $6.61$ & $4.78$\zstar & $5.76$\zstar \\
      $T_0(\rho_{\rm ft})$& $[{\rm keV}]$ &  N/A  & $0.18$ & $0.39$ & $0.18$ & $0.39$ & $0.26$\zstar & $0.33$\zstar \\
      $\beta(\rho_{\rm ft})$ & $[\%]$ &    N/A & $0.07$ & $0.29$ & $0.07$ & $0.29$ & $0.15$ & $0.22$ \\
      $\rho_*(\rho_{\rm ft})$& $[\%]$ &   N/A& $0.07$ & $0.11$ &  $0.07$ & $0.11$ & $0.09$\zstar & $0.10$\zstar \\
          $\nu_*(\rho_{\rm ft})$& [{\sc gene} units] &N/A&$0.04$&$0.02$&$0.04$&$0.02$& $0.02$\zstar & $0.02$\zstar\\
  \end{tabular}
  \caption{Table with parameters of various geometries considered in this Letter. Since only the pressure is being interpolated in LLH and LHH there is freedom in the choice of density and temperature, with the chosen values (and derived quantities) indicated with a superscript $*$.}
  \label{tab:vals}
  \end{center}
\end{table}

\section{The effect of current in the vicinity of a separatrix} \label{app: app separatrix}
\cite{bishop1986stability,bishop1987bifurcated,bishop1988theory} demonstrated that the first and second stability regions for ideal-MHD ballooning modes can coalesce if a sufficiently strong toroidal current is present on the investigated flux surface, if the latter is located near the separatrix. This stabilisation is a consequence of the change in local magnetic shear and was invoked by Bishop to argue for the existence of bifurcated temperature profiles akin to those observed in L- and H-mode. This Letter appeals to a related mechanism in gyrokinetics, but in the main body of the paper the effect of a separatrix on the magnetic equilibrium was not considered. Since a separatrix leads to infinite global magnetic shear, one may be concerned that the stabilising effect of the bootstrap current may be ineffective in a plasma with a poloidal divertor. As shown in this appendix, however, the bootstrap current can significantly reduce the magnetic shear, even close to the separatrix. 
\par
Our starting point is the following expression for the safety factor in an axisymmetric magnetic field ${\bf B} = F(\psi_p) \nabla \varphi + \nabla \varphi \times \nabla \psi_p$, 
\begin{equation}
    q(\psi_p) = \frac{F(\psi_p) V'(\psi_p)}{4 \pi^2} \left\langle \frac{1}{R^2} \right\rangle_{\rm fs},
\end{equation}
where $R$ denotes the major radius and $V(\psi_p)$ the volume enclosed by the flux surface $\psi_p$ \citep{HS}. As the separatrix is approached, the flux-surface average becomes dominated by the contribution from immediate vicinity of the X-point, and the magnetic shear becomes proportional to 
    \begin{equation}
    q'(\psi_p) = \frac{B_{\varphi, {\rm X}} V''(\psi_p)}{4 \pi^2 R_{\rm X}}, 
\end{equation}
where a subscript X indicates that the quantity in question is to be evaluated at the X-point. In order to calculate $V''(\psi_p)$, it is useful to note the identity \citep[Sec.~2.3]{helander2014theory},
\begin{equation}
    \frac{V''(\psi_p)}{V'(\psi_p)} = \left \langle \nabla \cdot \left( \frac{\nabla \psi_p}{|\nabla \psi_p|^2} \right) \right \rangle_{\rm fs}.
    \label{eq: magnetic hill}
\end{equation}
The Grad--Shafranov equation can be used to rewrite this expression in terms of toroidal current density ($j_\varphi$) and the curvature ($\kappa_p$) of the field lines in the poloidal plane, giving\footnote{Here, we have used $\nabla (|\nabla \psi_p|) \cdot \nabla \psi_p/|\nabla \psi_p|^3 = \boldsymbol{\hat{n}} \cdot (\nabla \nabla\psi_p) \cdot \boldsymbol{\hat{n}}/|\nabla \psi_p|^2$, where $\boldsymbol{\hat{n}} = \nabla \psi_p / |\nabla \psi_p|$. Substitute this term by one involving the Laplacian $\nabla^2 \psi_p= \mathrm{Tr} \left[ \nabla\nabla \psi_p \right]$, expressed using an orthonormal normal-tangential-toroidal basis $\{\boldsymbol{\hat{n}},\boldsymbol{\hat{\tau}},\boldsymbol{\hat{\varphi}}\}$. The toroidal component can be simplified as $\boldsymbol{\hat{\varphi}} \cdot (\nabla \nabla \psi_p) \cdot \boldsymbol{\hat{\varphi}} = \partial_R \psi_p/R$, and one finds $\nabla \cdot (\nabla \psi_p/|\nabla \psi_p|^2) = \left(-\Delta^* \psi_p + 2 \boldsymbol{\hat{\tau}} \cdot[ \nabla \nabla \psi_p ] \cdot \boldsymbol{\hat{\tau}}\right)/|\nabla \psi_p|^2$, where $\Delta^*$ is the Grad--Shafranov operator. Finally, realise that the final term relates to the poloidal curvature via $\boldsymbol{\hat{\tau}} \cdot( \nabla \nabla \psi_p ) \cdot \boldsymbol{\hat{\tau}} = \boldsymbol{\hat{\tau}} \cdot \nabla( \boldsymbol{\hat{\tau}} \cdot \nabla \psi_p) - (\boldsymbol{\hat{\tau}} \cdot \nabla \boldsymbol{\hat{\tau}}) \cdot \nabla \psi_p = \kappa_p |\nabla \psi_p|$.}
\begin{equation}
    \frac{V''(\psi_p)}{V'(\psi_p)} = \left \langle \frac{2 \kappa_p}{|\nabla \psi_p|} - \frac{\mu_0 R j_\varphi}{|\nabla \psi_p|^2} \right \rangle_{\rm fs}.
    \label{eq: magnetic hill}
\end{equation}
\par 

Since the magnetic field is purely toroidal at the X-point, the safety factor is infinite on the separatrix and the magnetic shear is unbounded in its vicinity. However, a realistic amount of bootstrap current can reverse the shear even quite close to the separatrix. In order to estimate how much current is required, it is helpful to write the magnetic field as
    $$ {\bf B} = {\bf B}_0 + \delta {\bf B}, $$
where $\delta {\bf B}$ denotes the modification, due to the bootstrap current at the plasma edge, of the ``background'' magnetic field ${\bf B}_0$ . We assume $|\delta {\bf B}| \ll B_0$ but $\nabla \delta {\bf B} \sim \nabla {\bf B}_0$, so that the density of the bootstrap current creating $\delta {\bf B}$ is allowed to be comparable to that of the main plasma current but is localised to a thin layer. The magnetic shear can thus be perturbed by the bootstrap current significantly, but the shape of the flux surfaces and the safety factor are not. (For a recent analysis of cases where the magnetic geometry is significantly perturbed, see \citealt{zheng2023asymptotic}). 

With these assumptions, the magnetic field lines in the vicinity of the X-point have approximately the same shape as if the current density were equal to a constant $j_{\rm X}$, and the poloidal flux-function can be expanded around the X-point in locally Cartesian coordinates $(u,v)$, which are orthogonal in the $(R,Z)$-plane, satisfying $\psi_{p,{\rm X}}-\psi_{p}\approx(\mathcal{U}^2u^2-\mathcal{V}^2v^2)/2$, where $\mathcal{U}$ and $\mathcal{V}$ are set by $\mathcal{U}^2-\mathcal{V}^2=-\mu_0R_{\rm X}j_{\rm X}$. A distance $\Delta\psi_p\equiv\delta^2/2$ from the X-point, the flux surfaces and the magnitude of the poloidal-flux gradient may be parametrised as
\begin{equation}
  u=\frac{\delta}{\mathcal{U}}\cosh t,\qquad
  v=\frac{\delta}{\mathcal{V}}\sinh t,\qquad
  |\nabla\psi_p|=\delta\sqrt{\mathcal{U}^2\cosh^2t+\mathcal{V}^2\sinh^2t}.
  \label{eq:param}
\end{equation}
The integrals in \eqref{eq: magnetic hill} are then dominated by the region around the X-point and may be calculated analytically as follows. The flux-surface average can be written as $\langle \ldots \rangle = (\oint \dd{\ell_p} \ldots/B_{p})/(\oint \dd{\ell_p} /B_{p})$, with $\ell_p$ being the arc-length in the poloidal plane and $B_p$ being the poloidal magnetic field. Note that to leading order $B_p \approx |\nabla \psi_p|/R_{\rm X}$, furthermore $\mathrm{d}\ell_p / \mathrm{d} t = |\nabla \psi_p|/\mathcal{UV}$, and finally $\kappa_p=(\mathcal{UV}\delta)^2/ |\nabla \psi_p|^3$. Filling in these results yields the following integrals entering $V''$,
\refstepcounter{equation}
\[
  \oint \frac{\dd{\ell_p} }{B_p} \frac{\kappa_p}{|\nabla \psi_p|} \approx \frac{R_{\rm X}\mathcal{UV}}{\delta^2} \oint \frac{\dd{t}\delta^4}{|\nabla \psi_p|^4}, \qquad \oint \frac{\dd{\ell_p} }{B_p} \frac{R j_{\rm BS}}{|\nabla \psi_p|^2} \approx \frac{R_{\rm X}^2 j_{\rm BS}}{\mathcal{UV} \delta^2} \oint  \frac{\dd{t}\delta^2}{|\nabla \psi_p|^2},
  \eqno{(\theequation{\mathit{a},\mathit{b}})}\label{eq: integrals}
\]
so that the region around the X-point dominates as $\mathcal{O}(\delta^{-2})$, and we may let $t \in (-\infty,\infty)$ since the tail vanishes exponentially. The remaining integrals can be evaluated in closed form and yield a condition on the amplitude of $j_{\rm BS}$ for shear reversal,
\begin{equation}
  R_{\rm X}\mu_0\big(j^{\mathrm{crit}}_{\rm BS}-j_{\rm X}\big)\approx\frac{\mathcal{UV}}{\arctan(\mathcal{V}/\mathcal{U})}.
  \label{eq:crit}
\end{equation}
Hence we see that a sharply peaked bootstrap current due to the pedestal can reverse the magnetic shear over a finite layer approaching to the X-point, despite the divergence of $\hat{s}$ even nearer to it, and this shear reversal is expected to be stabilising for curvature-driven modes in the manner discussed in the main text. \par
Finally, we may estimate how large this current density needs to be comparatively to the total current density (i.e. the plasma current over the area). Let the typical poloidal edge field value be $B_{p,a}$, which relates to the plasma current via Amp\`ere as $\overline{\jmath} = (B_{p,a}/\mu_0) (\mathcal{P} / \mathcal{A})$, where $\mathcal{P}$ and $\mathcal{A}$ are the perimeter length and area of the last-closed flux surface in the poloidal plane respectively. This yields
\begin{equation}
\begin{aligned}
  \frac{j^{\mathrm{crit}}_{\rm BS}-j_{\rm X}}{\overline{\jmath}}&\approx\frac{1}{\arctan(\mathcal{V}/\mathcal{U})} \frac{\mathcal{A}\mathcal{UV}}{B_{p,a}\mathcal{P} R_{\rm X}} \\ 
  &\approx \frac{a_{\rm h}/L_{\rm X}}{\arctan(\mathcal{V}/\mathcal{U})}
\end{aligned}
  \label{eq:crit}
\end{equation}
where we recognise that area-over-perimeter is the hydraulic radius $a_{\rm h}$ (which for circular cross-sections reduces to $a/2$). The remaining length-scale $L_{\rm X}$ can be interpreted by noting that the poloidal field (derived from the linearised X-point) reaches its typical value $B_{p,a}$ when $\mathcal{U}^4u^2 +\mathcal{V}^4v^2=R_{\rm X}^2B_{p,a}^2$, which is an ellipse centred on the X-point whose geometric radius is $L_{\rm X} = R_{\rm X} B_{p,a}/\mathcal{UV}$, it thus measures the extent of the null. Assuming no current at the X-point, and $a_{\rm h}\approx L_{\rm X}$, we find $j_{\rm BS}^{\rm crit}\approx (4/\pi)\, \overline{\jmath}$, very close to the criterion for shear reversal from Eq. \eqref{eq:shear-relation}.
It should be said that we everywhere assume the gyroradius remains much smaller than the distance to the X-point, thus neglecting ion-orbit loss effects \citep{ku2004property}.}

\bibliographystyle{jpp}
\bibliography{AAEbib.bib}

\end{document}